\documentclass{iaus}
\usepackage{graphicx}

\title[IAU 249.~~Planet-induced X-ray emission: HD~179949]
{First observation of planet-induced  \\
X-ray emission: The system HD~179949}

\author[S. H. Saar et al.]   %% give here short author list %%
{S. H. Saar$^1$, M. Cuntz$^2$, V. L. Kashyap$^1$ and J. C. Hall$^3$}

\affiliation{
$^1$Harvard-Smithsonian Center for Astrophysics, \\
Cambridge, MA 02138, USA \\
email: {\tt saar@cfa.harvard.edu, kashyap@cfa.harvard.edu} \\
[\affilskip]
$^2$Department of Physics, University of Texas at Arlington, \\
Arlington, TX 76019-0059, USA \\
email: {\tt cuntz@uta.edu} \\
[\affilskip]
$^3$Lowell Observatory, Flagstaff, AZ 86001, USA \\
email: {\tt jch@lowell.edu} \\
}

\pubyear{2008}
\volume{xxx}  %% insert here IAU Symposium No.
\pagerange{119--126}
\jname{Title of your IAU Symposium}
\editors{A.C. Editor, B.D. Editor \& C.E. Editor, eds.}
\begin{document}

\maketitle

\begin{abstract}
We present the first observation of planet-induced stellar X-ray activity,
identified for the HD 179949 system, using Chandra / ACIS-S.  The HD~179949
system consists of a close-in giant planet orbiting an F9~V star.
Previous ground-based observations already showed enhancements in Ca~II K
in phase with the planetary orbit.  We find an $\sim$30\% increase in the
X-ray flux over quiescent levels coincident with the phase of the Ca~II
enhancements.  There is also a trend for the emission to be hotter at
increased fluxes, confirmed by modeling, showing the enhancement at
$\sim$1 keV compared to $\sim$0.4 keV for the background star.
\keywords{Planetary systems, stars: activity, stars: coronae,
stars: individual (HD~179949), stars: late-type, stars: magnetic fields}

\end{abstract}

\firstsection % if your document starts with a section,

\section{Introduction}

Planets have been discovered around a large number of stars, mostly by the
cycle Doppler shift of their photospheric lines.  Most of these planets
have been found around F, G, and K-type main-sequence stars.  Moreover,
about 20\% of these planets are at an orbital distance of 0.1~AU or less
(e.g., \cite[Butler et al. 2006]{butl06}), commonly referred to as
close-in extrasolar giant planets (CEGPs).  An interesting question is
whether the CEGPs have any effect on the atmosphere of their parent star.
Using observed star-planet systems as a basis, \cite[Cuntz et al. (2000)]{cunt00}
were first to propose that CEPGs can increase chromospheric and coronal activity.
This effect was thereafter identified through high-quality data, obtained by
Ca~II K observations by \cite[Shkolnik et al. (2003)]{shko03} for five stars
(i.e., HD~179949, HD~209458, $\tau$~Boo, 51~Peg, $\upsilon$~And).  The
observations indicated unambigious star-planet activity enhancement
in the HD~179949 system.  Subsequent results were given by
\cite[Shkolnik et al. (2005)]{shko05}.

The enhancement was found to be phased with the planet orbital period
($P_{\rm orb} \simeq 3.09$~d), and not the (poorly known) stellar rotation
period, $P_{\rm rot}$, estimated to lie between 7 and 11~d.  This clearly
implies that the Ca~II emission enhancement is caused by some form of
star-planet interaction.  The peak excess amounts to $\sim$0.7\%
in the continuum-normalized K line core strength (which translates to
a $\sim$12\% increase in the [basal-subtracted] chromospheric flux; see
\cite[Saar et al. 2008]{saar08}), and is shifted in phase by 
$\Delta\phi_{\rm orb} \sim 0.18$ from the planet's inferior conjunction.
The presence of only one emission peak per $P_{\rm orb}$ seems to rule out
a tidal interaction that would result in a period of $P_{\rm orb}/2$; i.e.,
one peak per tidal bulge.  A variety of models have been proposed to explain
the observations, which indicate that planet-induced stellar activity
enhancements can be an important probe of (1) the close-in stellar magnetic
field structure, (2) stellar wind properties, and/or (3) the planetary
magnetosphere
(\cite[Saar et al. 2004]{saar04},
 \cite[Grie{\ss}meier et al. 2004]{grie04},
 \cite[Preusse et al. 2005]{preu05},
 \cite[McIvor et al. 2006]{mciv06},
 \cite[Zarka 2007]{zark07},
 \cite[Cranmer \& Saar (2007)]{cran07}).
In the following, we report the first
observation of planet-induced X-ray emission, found in the exoplanetary
system HD~179949.

%%% *** Fig.1
%%%%%%%%%%%%%%%%%%%%%%%%%%%%%%%%%%%%%%%%%%%%%%%%%%%%%%%%%%%%%%%%%
\begin{figure}
\begin{center}
\includegraphics[width=3.2in]{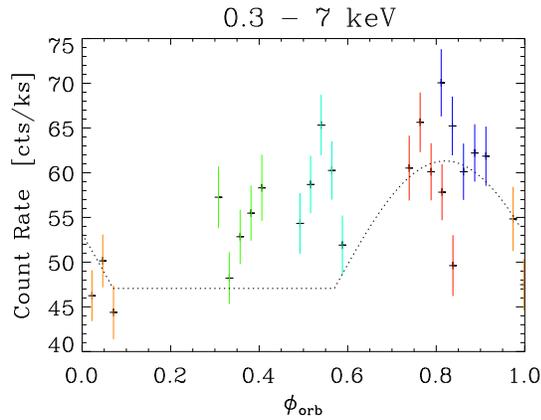}
\caption{
Background subtracted ACIS-S count rate ($0.3-7$ keV; 6 ksec bins)
as a function of the planet orbital phase $\phi_{\rm orb}$ for various
$P_{\rm rot}$.  Observations were taken in 30 ksec segments separated
by 1, 8, 9, and $10 \times P_{\rm orb}$ (orange, green, light \& dark blue,
respectively) from the first (red).  We also show the best fitting Ca~II
H+K model following \cite[Shkolnik et al. (2003)]{shko03} (dotted).
}
\end{center}
\end{figure}
%%%%%%%%%%%%%%%%%%%%%%%%%%%%%%%%%%%%%%%%%%%%%%%%%%%%%%%%%%%%%%%%%

%%% *** Fig.2
%%%%%%%%%%%%%%%%%%%%%%%%%%%%%%%%%%%%%%%%%%%%%%%%%%%%%%%%%%%%%%%%%
\begin{figure}
\begin{center}
\includegraphics[width=3.2in]{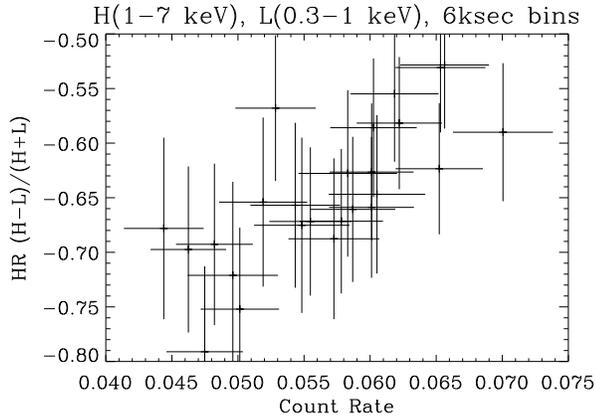}
\caption{
Hardness ratio between high ($1-6$ keV) and low ($0.3-0.6$ keV) energies
as a function of the total count rate.
}
\end{center}
\end{figure}
%%%%%%%%%%%%%%%%%%%%%%%%%%%%%%%%%%%%%%%%%%%%%%%%%%%%%%%%%%%%%%%%%

\section{Observations and Interpretation}

To reduce complications due to varying stellar activity, the observations
were taken at nearly the same stellar rotational phase $\phi_{\rm rot}$
and within 3~$P_{\rm rot}$, minimizing possible spatial and temporal changes,
respectively; see \cite[Saar et al. (2008)]{saar08} for details.
For 7.5~d $\le P_{\rm rot} \le$ 10~d, the phase span
$\Delta\phi_{\rm rot} \le 0.3$.  Plots versus $\phi_{\rm rot}$
(not given here) show either scatter ($P_{\rm rot} = 8-9$~d)
or sharp changes over multiple orbits difficult to explain with
rotational modulation.  A Lomb-Scargle periodiogram analysis yields
$P = 3.289$~d (false alarm probability = $2 \times 10^{-11}$) with
no other $P$ consistent with possible $P_{\rm rot}$ values; thus a
planet-related origin for any variation is preferred.

The data phased to $\phi_{\rm orb}$ show a minimum around
$\phi \sim 0.0 - 0.1$ and a gradual rise to maximum around
$\phi \sim 0.7 - 0.9$ (Fig. 1).  Nearly all the variation is at
high energies ($>1$ keV); additionally, the hardness ratio correlates well
with total count rate (Fig. 2). The variation at high energies is also
not smooth.  Thus the ``planet effect" produces hot, fluctuating (flare-like?)
variability.  Contemporaneous Ca II H+K data from the Lowell Observatory
Solar-Stellar Spectrograph is consistent (within the large errors)
with variations seen in \cite[Shkolnik et al. (2005)]{shko05}.
The best fit to a scaled version of their H+K model, however (fixing the minimum
to the average flux in $\phi \sim 0.0 - 0.1$ and the emission peak phase shift
$\Delta\phi_{\rm orb} = 0.18$), shows significant unmodeled excess flux in the range
$\phi \sim 0.4 - 0.6$ (Fig. 1).  

\section{Summary}

We have detected, in the HD 179949 exoplanet system, the X-ray counterpart
to the excess Ca~II H+K emission previously obtained that is phased to
the planet's orbit.  The following results are forwarded by the observations:

\begin{enumerate}

\item
Peak X-ray enhancement ($0.3-7$ keV) over the background is $\sim$6 times that 
seen in Ca~II H+K and shows a similar phase shift from inferior
conjunction, which is $\Delta\phi_{\rm orb} \sim 0.18$.

\item
Thermal plasma models indicate the background has $T \sim 0.4$~keV,
consistent with the corona of a modestly active star.  The component
responsible for the variability, and associated with the planet,
is hotter with $T \sim 1$~keV and ${\rm EM}_{\rm hot}/{\rm EM}_{\rm cool}
\sim 0.3$.

\item
There is significant additional excess flux around $\phi \sim 0.4-0.6$,
which is also hot ($T \sim 1$~keV), but does not follow the variation seen
in Ca~II H+K.  The source of this emission is unclear; it may come from interaction
with a second loop (possibly rooted at high stellar latitude) yielding a 
larger $\Delta\phi$, or it may be emission from the planet's
magnetosphere itself (most visible near $\phi = 0.5$).  These different
scenarios will be explored in future studies
(\cite[Saar et al. 2008]{saar08}).

\end{enumerate}

Additional observations and modeling of this phenomenon are needed to obtain
further insight into the dominant physical processes.


\begin{thebibliography}{}

\bibitem[Butler et al. (2006)]{butl06}
{Butler, R.P., et al.} 2006, \textit{ApJ}, 646, 505

\bibitem[Cranmer \& Saar (2007)]{cran07}
{Cranmer, S.R., \& Saar, S.H.}, astro-ph/0702530

\bibitem[Cuntz et al. (2000)]{cunt00}
{Cuntz, M., Saar, S.H., \& Musielak, Z.E.} 2000, \textit{ApJ} (Letters), 533, L151

\bibitem[Grie{\ss}meier et al. (2004)]{grie04}
{Grie{\ss}meier, J.-M., et al.} 2004, \textit{A\&A}, 425, 753

\bibitem[McIvor et al. (2006)]{mciv06}
{McIvor, T., Jardine, M., \& Holzwarth, V.} 2006, \textit{MNRAS}, 367, L1

\bibitem[Preusse et al. (2005)]{preu05}
{Preusse, S., Kopp, A., B\"uchner, J., \& Motschmann, U.} 2005, \textit{A\&A},
434, 1191

\bibitem[Saar et al. (2004)]{saar04}
{Saar, S.H., Cuntz, M., \& Shkolnik, E.} 2004, in: A.K. Dupree \& A.O. Benz
(eds.), \textit{Stars as Suns: Activity, Evolution and Planets},
IAU Symp. 219 (San Francisco: ASP), p. 355

\bibitem[Saar et al. (2008)]{saar08}
{Saar, S.H., Kashyap, V.L., Cuntz, M., Shkolnik, E., \& Hall, J.C.} 2008,
\textit{ApJ}, in preparation

\bibitem[Shkolnik et al. (2003)]{shko03}
{Shkolnik, E., Walker, G.A.H., \& Bohlender, D.A.} 2003, \textit{ApJ}, 597, 1092;
Erratum 609, 1197 [2004]

\bibitem[Shkolnik et al. (2005)]{shko05}
{Shkolnik, E., Walker, G.A.H., Bohlender, D.A., Gu, P.-G., \& K\"urster, M.}
2005, \textit{ApJ}, 622, 1075

\bibitem[Zarka (2007)]{zark07}
{Zarka, P.} 2007, \textit{Planetary and Space Science}, 55 (5), 598

\end{thebibliography}
\end{document}